\documentclass[sigconf]{acmart}
\AtBeginDocument{%
  }

\usepackage{graphicx}
\usepackage{booktabs}
\usepackage{enumitem}
\usepackage{diagbox} 
\usepackage{cases}
\usepackage{multirow}
\usepackage{multicol}
\usepackage{adjustbox}
\usepackage{amsfonts}

\usepackage{subfig}
\usepackage{xcolor}

\setcopyright{acmlicensed}
\copyrightyear{2026}
\acmYear{2026}
\acmDOI{XXXXXXX.XXXXXXX}
\acmConference[ICMR'26]{ACM Conference}{June 16 - June 19, 2026}{Amsterdam, The Netherlands}

\acmISBN{978-1-4503-XXXX-X/2018/06}

\begin{document}

\title{Zero-Effort Image-to-Music Generation: An Interpretable RAG-based VLM Approach}

\author{Zijian Zhao}
\authornote{Corresponding Author}
\affiliation{%
  \institution{The Hong Kong University of Science and Technology}
  \city{Hong Kong}
  \country{China}}
   \email{zzhaock@connect.ust.hk}
    \orcid{0000-0002-3326-9650}

\author{Dian Jin}
\affiliation{%
  \institution{The Hong Kong Polytechnic University}
  \city{Hong Kong}
  \country{China}}
   \email{diann.jin@connect.polyu.hk}

\author{Zijing Zhou}
\affiliation{%
  \institution{The University of Hong Kong}
  \city{Hong Kong}
  \country{China}}
   \email{zhouzj18@lzu.edu.cn}

\renewcommand{\shortauthors}{Zhao et al.}

\begin{abstract}
  Recently, Image-to-Music (I2M) generation has garnered significant attention, with potential applications in fields such as gaming, advertising, and multi-modal art creation. However, due to the ambiguous and subjective nature of I2M tasks, most end-to-end methods lack interpretability, leaving users puzzled about the generation results. Even methods based on emotion mapping face controversy, as emotion represents only a singular aspect of art. Additionally, most learning-based methods require substantial computational resources and large datasets for training, hindering accessibility for common users. To address these challenges, we propose the first Vision Language Model (VLM)-based I2M framework that offers high interpretability and low computational cost. Specifically, we utilize ABC notation to bridge the text and music modalities, enabling the VLM to generate music using natural language. We then apply multi-modal Retrieval-Augmented Generation (RAG) and self-refinement techniques to allow the VLM to produce high-quality music without external training. Furthermore, we leverage the generated motivations in text and the attention maps from the VLM to provide explanations for the generated results in both text and image modalities. To validate our method, we conduct both human studies and machine evaluations, where our method outperforms others in terms of music quality and music-image consistency, indicating promising results. Our code is available at \url{https://github.com/RS2002/Image2Music}.
\end{abstract}

\ccsdesc[500]{Applied computing~Sound and music computing}
\ccsdesc[300]{Computing methodologies~Artificial intelligence}

\keywords{Image-to-Music Generation (I2M), Vision Language Model (VLM), Retrieval-Augmented Generation (RAG), Music Information Retrieval (MIR), Symbolic Music, Multi-Modal, Interpretability}


\maketitle

\section{Introduction}

In recent years, multi-modal music generation has made significant strides, encompassing diverse areas such as video background music generation \cite{di2021video} and natural language-controlled music generation \cite{melechovsky2024mustango}. Among these, Image-to-Music (I2M) presents a range of promising applications, including assisting visually impaired individuals in experiencing visual art, facilitating multi-modal art creation, developing Human-Computer Interaction (HCI) tools like music albums, and supporting therapeutic practices through Guided Imagery and Music (GIM) \cite{bonny2002music}.

Despite these advancements, the field faces several critical challenges: (i) The mapping from images to music is inherently ambiguous and subjective \cite{wang2023continuous}. Although various approaches, such as emotion-based mappings \cite{wang2023continuous,sergio2015generating,tan2020automated,kundu2024emotion,hisariya2024bridging}, have been proposed, they often lack comprehensiveness and rationality \cite{zhao2025automatic,mcdonald2022illuminating}. (ii) While there are a limited number of datasets available \cite{liu2023m}, assessing their quality remains challenging. The performance of models trained on these datasets can be significantly affected. Furthermore, even if a dataset is suitable for training an end-to-end model, the resulting system may lack interpretability, hindering users' understanding of the connections between images and generated music. (iii) Most I2M systems depend on complex model architectures \cite{chowdhury2024melfusion}, which require substantial training and computational resources, especially those that involve fine-tuning large pre-trained models \cite{wang2024multimodal,liu2024mumu,tian2025xmusic}.

To address these challenges, we propose a novel framework based on Vision Language Models (VLMs) that offers high interpretability without the need for additional training. Our approach leverages the image understanding and text generation capabilities of pre-trained VLMs by utilizing ABC music notation \cite{ji2023survey} to bridge natural language and symbolic music. By introducing a Multi-Modal Retrieval-Augmented Generation (RAG) mechanism, we equip the VLM with sufficient knowledge about music generation, thereby eliminating the need for external training. This framework not only provides a multi-modal explanation for the motivation behind music creation, where the textual explanation derives from the VLM output and the image explanation is based on a processed attention map within the VLM, but also enhances music quality through a novel self-reflection mechanism. This mechanism allows the VLM to refine its generated music according to various quality metrics. The contributions of this paper can be summarized as follows:

\begin{itemize}[noitemsep,leftmargin=*]
    \item We introduce a novel framework for the I2M task characterized by low computational cost and high interpretability. 
    Our method directly utilizes pre-trained VLMs to generate text-based music in ABC notation from images, while ensuring interpretability through two complementary forms: the linguistic explanations provided by the VLM’s textual output and the visual reasoning revealed in the attention map.

    \vspace{0.1cm}
    \item To enhance the quality of generated music, we present a multi-modal RAG method that supplies the VLM with external knowledge about music. Additionally, we implement a self-reflection generation process, employing an evaluator to guide the VLM in refining its musical outputs.
    
    \vspace{0.1cm}
    \item We validate our method through both human studies and machine evaluations, comparing it with previous I2M methods and conducting several ablation studies. The results show that our method achieves the highest scores in terms of both music quality and music-image consistency, demonstrating the effectiveness of our design.
\end{itemize}

\section{Methodology}

\begin{figure*}[htbp]
\centering 
\includegraphics[width=\textwidth]{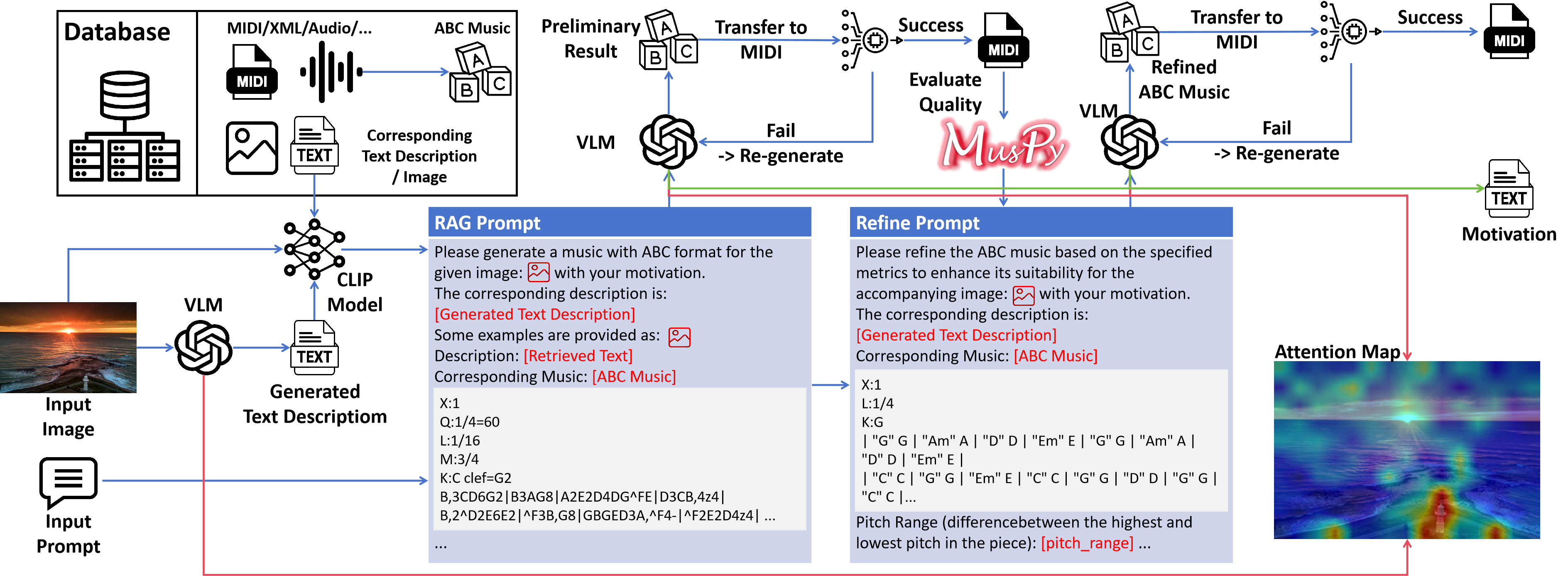}
\caption{Overall Workflow of Proposed Framework}
\label{fig:main}
\end{figure*}

The workflow of our method is illustrated in Fig. \ref{fig:main}, comprising three main phases: (i) primary music generation with the multi-modal RAG; (ii) music refinement using a model-based evaluator; and (iii) explanation generation with text output and an image attention map. In this section, we will provide a detailed introduction to the process and underlying motivation for each step.

\subsection{Music Generation with Multi-Modal RAG}
Large models have demonstrated promising generalization capabilities, largely due to extensive training datasets and pre-training mechanisms. However, two main challenges arise when directly using VLMs or Large Language Models (LLMs) for music generation.

First, these models struggle to directly generate audio files. A common solution involves using symbolic music as a bridge between the language and audio modalities. Among various music representation formats, the ABC notation (illustrated in the gray boxes in Fig. \ref{fig:main}) stands out. It utilizes music papers (A-G) and numbers to represent pitch and duration, making it easily understandable for both humans and machines, in contrast to formats like MIDI or MusicXML.

Second, the training corpus of many current large models contains limited information about music, particularly regarding music generation. This limitation makes it challenging for them to produce ABC music directly. To develop domain-specific models from large models \cite{chen2025overview}, fine-tuning and RAG are the two predominant approaches. Fine-tuning often requires substantial training resources and data. In contrast, RAG offers a plug-and-play method that demonstrates high flexibility and efficiency with promising results. Below, we formally describe our RAG-based music generation process.

Consider a database containing pieces of music along with their descriptions (which may be text or images). For VLM understanding, we first convert the music into ABC notation, represented as $M=[m_1,m_2,\ldots,m_n]$ where $n$ denotes the number of music pieces. To utilize RAG, we employ a trained Contrastive Language-Image Pre-training (CLIP) model \cite{radford2021learning} to encode the descriptions $D=[d_1,d_2,\ldots,d_n]$ into embeddings $E=[e_1,e_2,\ldots,e_n] \in \mathbb{R}^{n,h}$, where $h$ is the dimension of the hidden features. Next, given an input image $I$, we first use the VLM to generate a text description $T$ and then encode it with the CLIP to obtain embeddings $\mathcal{E}_I$ and $\mathcal{E}_T$, respectively. We can then calculate the similarity between the input image and the music in the database using cosine similarity among the embeddings:
\begin{equation} 
S(\mathcal{E},e_j) = \frac{\mathcal{E} \cdot e_j}{||\mathcal{E}||\cdot||e_j||} \ ,
\label{eq:sim}
\end{equation}
where $||\cdot||$ denotes the Euclidean norm. Inspired by \cite{chen2025each}, we select the top $k$ music pieces with the highest similarity to both the text description $T$ and the original image $I$. We then incorporate these music pieces along with their descriptions into the prompt and ask the VLM to generate corresponding music for the input image $I$ using these references.

\subsection{Music Refinement with Quality Evaluator}
Multi-round generation has become a popular method in generative models for quality refinement. To achieve this, we propose two approaches for VLM-based music generation. First, we utilize a parser to convert the generated ABC music into MIDI files. If there are any grammatical errors, the parser will report them, prompting the VLM to regenerate the music \cite{cao2025towards}. Second, we introduce an evaluator-based method that allows the VLM to self-reflect and refine the generated music. Specifically, we employ MusPy \cite{dong2020muspy} to assess the generated MIDI music using several common metrics, including Pitch Range (PR), Number of Pitches Used (NPU), Number of Pitch Classes Used (NPCU), Polyphony, Scale Consistency (SC), Pitch Entropy (PE), Pitch Class Entropy (PCE), and Empty Beat Rate (EBR) \cite{yang2020evaluation}. These metrics are style-independent: for example, higher entropy is suited for serialist compositions that aim for atonal equality, while lower entropy benefits pop music, where predictable pitch centers create catchy melodies. As a result, we provide the evaluated metric results, along with detailed explanations, back to the VLM. This enables the model to refine the music according to the given conditions (image and text description) based on its own judgment.

\subsection{Multi-Modal Explanation Generation}
To address the low interpretability problem of most current music generation models, we propose two approaches based on text and image modalities, respectively. For the text modality, we employ a straightforward method by directly asking the VLM to provide the motivation behind its creation using text. For the image modality, we utilize attention maps to analyze which parts of the input image the model focuses on when generating music, which has been widely used in various studies \cite{zhao2024mining}. For example, for a specific part of the generated text $t$, we can obtain the attention map to the image, denoted as $A \in \mathbb{R}^{l,h,|I|,|t|}$, where $l$ is the number of layers, $h$ is the number of heads, and $|I|$ and $|t|$ represent the token counts for the input image $I$ and the generated text $t$, respectively. We then calculate the average attention to the image as follows:
\begin{equation} 
\hat{A} = \mathrm{mean}(A, \mathrm{dim}=0,1,3) \in \mathbb{R}^{|I|} \ .
\label{eq:attn}
\end{equation}
This allows us to identify which parts of the image the model primarily utilizes when generating each track or segment of music. However, this method requires tracking the entire attention matrix during inference, which leads to high GPU utilization. In practice, this aspect should be considered based on the actual device capacity and the scale of the VLM used.

\begin{table*}[htbp]
\caption{Machine Evaluation by SongEval \cite{yao2025songeval}: The \textbf{bold} text indicates the best result, while \underline{underlined} text represents the second-best result. This formatting will be consistent in the following tables.}
\centering
    \begin{tabular}{l@{\quad}|cccccc}
        \toprule
        \textbf{\diagbox{Methods}{Metrics}} & \textbf{Coherence}  &  \textbf{Musicality} & \textbf{Memorability}  &  \textbf{Clarity}  &  \textbf{Naturalness} &  \textbf{Average}  \\
        \midrule
        \textbf{Synesthesia} \cite{tan2020automated} & 2.31 & 2.38 & 2.10 & 2.15 & 2.10 & 2.21 \\
        \textbf{Mozart's Touch} \cite{li2025mozart}  & \underline{2.84} & \underline{2.82} & \textbf{2.66} & \underline{2.63} & \underline{2.59} & \underline{2.71} \\
        \midrule
        \textbf{Ours} & \textbf{2.90} & \textbf{2.84} & \underline{2.60} & \textbf{2.66} & \textbf{2.60} & \textbf{2.72} \\
        \textbf{w/o RAG} & 2.78 & 2.63 & 2.54 & 2.49 & 2.48 & 2.58 \\
        \textbf{w/o Refinement} & 2.81 & 2.79 & 2.56 & 2.63 & 2.57 & 2.67 \\
        \textbf{w/o RAG, Refinement} & 2.76 & 2.62 & 2.52 & 2.46 & 2.41 & 2.56 \\
        \bottomrule
    \end{tabular}
    \label{tab:songeval}
\end{table*}

\begin{table*}[htbp]
\caption{Machine Evaluation by VLM judge}
\centering
    \begin{adjustbox}{width=1.00\textwidth}
    \begin{tabular}{l@{\quad}|cccccc|cccc}
        \toprule
        \multirow{2}{*}{\textbf{\diagbox{Methods}{Metrics}}} & \multicolumn{6}{c|}{\textbf{Music Quality}} & \multicolumn{4}{c}{\textbf{Music-Image Consistency}} \\
        & \textbf{Overall} & \textbf{Melody} & \textbf{Rhythm} & \textbf{Authenticity} & \textbf{Harmony} & \textbf{Average}  & \textbf{Overall} & \textbf{Semantics} & \textbf{Emotion} & \textbf{Average}  \\
        \midrule
        \textbf{Synesthesia} \cite{tan2020automated} & 3.0	& 2.4	& 3.6	& 2.8	& 2.2 & 2.8	& 3.8	& 2.8	& 3.8 & 3.5 \\
        \textbf{Mozart's Touch} \cite{li2025mozart}  & 4.4	& 3.8	& \underline{4.6}	& 4.6	& 4.2 & 4.3	& 5.4	& 4.0	& 5.6 & 5.0  \\
        \midrule
        \textbf{Ours} & \textbf{5.2}	& \textbf{5.2}	& \textbf{5.2}	& \textbf{5.2}	& \textbf{5.0} &	\textbf{5.2} & \textbf{6.0} 	& \textbf{5.2}	& \textbf{6.0} & \textbf{5.7} \\
        \textbf{w/o RAG} & 4.8	& 4.4	& 4.4	& 4.8	& 4.4 & 4.6	& 5.6	& \underline{5.0}	& 5.4 & 5.3  \\
        \textbf{w/o Refinement} & \textbf{5.2}	& \underline{5.0}	& \underline{4.6}	& \underline{5.0}	& \textbf{5.0}	& \underline{5.0} & \underline{5.8}	& \underline{5.0}	& \underline{5.8} & \underline{5.5}  \\
        \textbf{w/o RAG, Refinement}  & 4.8	& 4.4	& \underline{4.6} &	\underline{5.0} 	& 4.4 & 4.6	& 5.4	& 4.8	& 5.4  & 5.2\\
        \bottomrule
    \end{tabular}
    \end{adjustbox}
    \label{tab:machine}
\end{table*}

\begin{table*}[htbp]
\caption{Human Evaluation Results}
\centering
    \begin{adjustbox}{width=1.00\textwidth}
    \begin{tabular}{l@{\quad}|cccccc|cccc}
        \toprule
        \multirow{2}{*}{\textbf{\diagbox{Methods}{Metrics}}} & \multicolumn{6}{c|}{\textbf{Music Quality}} & \multicolumn{4}{c}{\textbf{Music-Image Consistency}} \\
        & \textbf{Overall} & \textbf{Melody} & \textbf{Rhythm} & \textbf{Authenticity} & \textbf{Harmony} & \textbf{Average} & \textbf{Overall} & \textbf{Semantics} & \textbf{Emotion} & \textbf{Average} \\
        \midrule
        \textbf{Synesthesia} \cite{tan2020automated} & 3.65 &	3.52 &	\underline{4.10} &	\textbf{4.04}	& 3.47	& \underline{3.75} & 	2.80 &	2.75	& 2.83	& 2.79 \\
        \textbf{Mozart's Touch} \cite{li2025mozart}  & \underline{3.69}	& \underline{3.57}	& 4.06	& 3.53 & \underline{3.82}	& 3.73	& \underline{3.40}	& \underline{3.14}	& \underline{3.53} &	\underline{3.35} \\
        \textbf{Ours} & \textbf{4.02}	& \textbf{3.73}	& \textbf{4.22}	& \underline{3.68}	& \textbf{4.02}	& \textbf{3.93}	& \textbf{3.88}	& \textbf{3.58} & 	\textbf{4.01} &	\textbf{3.82} \\
        \bottomrule
    \end{tabular}
    \end{adjustbox}
    \label{tab:human}
\end{table*}


\section{Experiment}

\subsection{Experiment Setup}

To implement our framework, we utilize the latest open-source model, Keye-8B \cite{kwaikeyeteam2025kwaikeyevltechnicalreport}, as the Vision Language Model (VLM). For the multi-modal CLIP component, we employ LongCLIP \cite{zhang2024longclip}, and we use MidiCaps \cite{melechovsky2024midicaps} as the external database, which contains 168,385 MIDI music files paired with descriptive text captions. The experiments were conducted using the PyTorch framework \cite{paszke1912pytorch} on a server running Ubuntu 22.04.5 LTS, equipped with an Intel(R) Xeon(R) Gold 6133 CPU @ 2.50 GHz and one NVIDIA A100 GPU. 

To validate the proposed method, we aim to compare our approach with existing ones. However, as noted by \cite{tian2025xmusic}, there are very few open-source I2M methods available, particularly for symbolic music generation. Since we require music to be formatted in ABC notation for evaluation with the language model LLM, audio-based methods may introduce errors and biases. Consequently, we follow \cite{tian2025xmusic} in selecting the open-source Synesthesia \cite{tan2020automated} as our benchmark. This method utilizes ResNet18 \cite{he2016deep} for image encoding, which serves as a prompt for RNN \cite{RNN} or Transformer \cite{vaswani2017attention} to generate music. Moreover, to demonstrate the superior performance of our method, we include a comparison with the latest State-Of-The-Art (SOTA) approach, Mozart's Touch \cite{li2025mozart}. This method first employs a caption generation model for image description and then inputs the description into a music generation method with LLM refinement. For fairness, we align the settings and module selections with our method. Moreover, we conduct a series of ablation studies to explore the efficacy of our proposed multi-modal RAG method and music refinement mechanism. In the experiment, we adhere to the experimental design outlined by \cite{tan2020automated}, incorporating both machine and human evaluations using the testing images from \cite{you2016building}.

\subsection{Machine Evaluation}

In the machine evaluation, we first utilize the latest music evaluation benchmark method, SongEval \cite{yao2025songeval}, to assess music quality in terms of coherence, musicality, memorability, clarity, and naturalness. The results shown in Table \ref{tab:songeval} illustrate that our method achieves SOTA performance in most metrics. Our approach, alongside Mozart's Touch, significantly outperforms the conventional deep learning method Synesthesia, which can be attributed to the extensive knowledge and emerging capabilities of large models. Additionally, the ablation study reveals that without our proposed multi-modal RAG and refinement method, the model performance experiences a significant decrease. This decrement is more pronounced when eliminating the RAG, suggesting that the retrieved music provides valuable references for generation.

To further evaluate music quality and music-image consistency, we design an evaluation utilizing VLM as judges, which have been widely employed in similar research \cite{li2025generation}. Inspired by \cite{wang2025vision,li2025survey}, we consider the following metrics: (i) music quality level, including overall quality, melody, rhythm, authenticity, and harmony; (ii) music-image consistency, encompassing overall correspondence, semantic consistency, and emotional consistency. During the experiment, we ask the VLM to score each image-music pair across these metrics, where the score ranges from 1 to 7 (with higher scores indicating better evaluations) \cite{zhao2025automatic}. 

For the evaluation, we select the latest multi-modal LLM, Grok4 \cite{grok}, converting the music into ABC notation for input into Grok4. The experimental results are illustrated in Table \ref{tab:machine}, where our method shows a consistent advantage in both music quality and music-image consistency. In the ablation study, we notice that the RAG module has a significant influence on both terms, as the retrieved image-music pair not only provides a strong example for music generation but also helps the VLM better understand the relationship between image and music through the retrieved example.

\subsection{Human Evaluation}

In regard to human evaluation, we designed a questionnaire to gather participants' feedback, utilizing the same questions posed to Grok4 in the previous subsection. We recruited 31 respondents through social media platforms and live music venues, with the human evaluation spanning half a month. The participants included 13 males and 18 females aged 18–55 (predominantly 18–30), with the majority possessing backgrounds in music or visual arts at varying levels, while 13 participants reported no relevant experience. All participants reported normal hearing, normal or corrected-to-normal vision, and no history of color blindness or color weakness.

During the human evaluation, we omitted the ablation study since increasing the number of comparative methods complicates data collection and may discourage user participation; longer questionnaires can lead to fatigue and confusion among participants. The experimental results are presented in Table \ref{tab:human}, where, although the overall scores from humans are lower than those from the machine evaluations, the trends are similar. Our method received the highest score, followed by Mozart's Touch and Synesthesia in that order.


\section{Conclusion}

This paper proposes a novel framework for the image-to-music generation task, offering advantages in terms of low cost and high interpretability. Based on a trained VLM, we introduce a series of modules, including multi-modal RAG, self-refinement, and prompt engineering, to generate high-quality music without the need for external training. Additionally, we leverage the motivations generated by the VLM in text format and the attention maps from images to provide explanations for the generated results. To evaluate the proposed method, we conduct both human and machine evaluations. The results demonstrate that our method achieves promising performance in music quality and image-music consistency, suggesting an efficient design. In the future, it would be worthwhile to further explore the potential of the proposed method using more powerful VLMs or music domain-specific models. Additionally, the reinforcement learning post-training with LLM as a judge for reward signals is also a potential paradigm under the limited labeled dataset in this field.

\newpage
\bibliographystyle{ACM-Reference-Format}
\bibliography{sample-base}

\end{document}